# Roadmap for Emerging Materials for Spintronic Device Applications

Atsufumi Hirohata,[1] *Senior Member, IEEE*, Hiroaki Sukegawa,[2] Hideto Yanagihara,[3] Igor Žutić,[4] Takeshi Seki,[5] Shigemi Mizukami[6] and Raja Swaminathan[7]

[1] Department of Electronics, University of York, York YO10 5DD, UK
[2] Magnetic Materials Unit, National Institute for Materials Science, Tsukuba 305-0047, Japan
[3] Graduate School of Pure and Applied Sciences, University of Tsukuba, Tsukuba 305-8577, Japan
[4] Department of Physics, University at Buffalo, State University of New York, Buffalo, NY 14260, USA
[5] Institute for Materials Research, Tohoku University, Sendai 980-8577, Japan
[6] WPI Advanced Institute for Materials Research, Tohoku University, Sendai 980-8577, Japan
[7] Intel Corporation, Chandler, AZ 85226, USA

The Technical Committee of the IEEE Magnetics Society has selected 7 research topics to develop their roadmaps, where major developments should be listed alongside expected timelines; (i) hard disk drives, (ii) magnetic random access memories, (iii) domain-wall devices, (iv) permanent magnets, (v) sensors and actuators, (vi) magnetic materials and (vii) organic devices. Among them, magnetic materials for spintronic devices have been surveyed as the first exercise. In this roadmap exercise, we have targeted magnetic tunnel and spin-valve junctions as spintronic devices. These can be used for example as a cell for a magnetic random access memory and spin-torque oscillator in their vertical form as well as a spin transistor and a spin Hall device in their lateral form. In these devices, the critical role of magnetic materials is to inject spin-polarised electrons efficiently into a non-magnet. We have accordingly identified 2 key properties to be achieved by developing new magnetic materials for future spintronic devices:
(1) Half-metallicity at room temperature (RT);
(2) Perpendicular anisotropy in nano-scale devices at RT.
For the first property, 5 major magnetic materials are selected for their evaluation for future magnetic/spintronic device applications: Heusler alloys, ferrites, rutiles, perovskites and dilute magnetic semiconductors. These alloys have been reported or predicted to be half-metallic ferromagnets at RT. They possess a bandgap at the Fermi level $E_F$ only for its minority spins, achieving 100% spin polarisation at $E_F$. We have also evaluated $L1_0$-alloys and $D0_{22}$-Mn-alloys for the development of a perpendicularly anisotropic ferromagnet with large spin polarisation. We have listed several key milestones for each material on their functionality improvements, property achievements, device implementations and interdisciplinary applications within 35 years time scale. The individual analyses and projections are discussed in the following sections.

*Index Terms*—Magnetic materials, half-metallic ferromagnets, Magnetic anisotropy, Spintronics.

## I. Heusler Alloys

HEUSLER alloys are ternary alloys originally discovered by Heusler [1]. He demonstrated ferromagnetic behaviour in an alloy consisting of non-magnetic atoms, $Cu_2MnSn$. Since then, these alloys have been investigated due to their properties of shape-memory and thermal conductance. In 1983, de Groot *et al.* reported half-metallic ferromagnetism in one of the Heusler alloys, half-Heusler NiMnSb alloy [2]. A great deal of effort has been accordingly devoted to achieve the half-metallicity at RT using a Heusler alloy. In particular, Block *et al.* measured a large tunnelling magnetoresistance (TMR) in bulk full-Heusler $Co_2(Cr,Fe)Si$ alloy [3], followed by a similar measurement in a thin-film form [4].

Among these Heusler alloys, Co-based full-Heusler alloys are the most promising candidates to achieve the RT half-metallicity due to their high Curie temperature ($T_C \gg$ RT), good lattice matching with major substrates, large minority-spin bandgap ($\geq 0.4$ eV, see Fig. 1), and large magnetic moments in general [$\geq 4$ $\mu_B$ per formula unit (f.u.)] [5],[6]. The main obstacle to achieving the half-metallicity in the Heusler-alloy films is the vulnerability against the crystalline disorder, such as the atomic displacement, misfit dislocation and symmetry break in the vicinity of the surface of the films. For the full-Heusler alloys forming $X_2YZ$, where the X and Y atoms are transition metals, while Z is either a semiconductor or a non-magnetic metal, the unit cell of the ideal crystalline structure ($L2_1$ phase, see Fig. 2.1) consists of four face-centered cubic (fcc) sublattices. When the Y and Z atoms exchange their sites (Y-Z disorder) and eventually occupy their sites at random, the alloy transforms into the $B2$ phase. In addition, X-Y and X-Z disorder finally leads to the formation of the $A2$ phase. By increasing the disorder, the magnetic properties depart further from the half-metallicity.

Towards the RT half-metallicity, two milestones have been identified as listed below:

(m1.1) Demonstration of >100% giant magnetoresistance (GMR) ratio at RT;

(m1.2) Demonstration of >1,000% tunnelling magnetoresistance (TMR) ratio at RT.

Here, we have regarded these criteria using the MR as indicator of the half-metallicity at RT

Corresponding author: A. Hirohata (e-mail: atsufumi.hirohata@york.ac.uk). All the authors contributed equally.

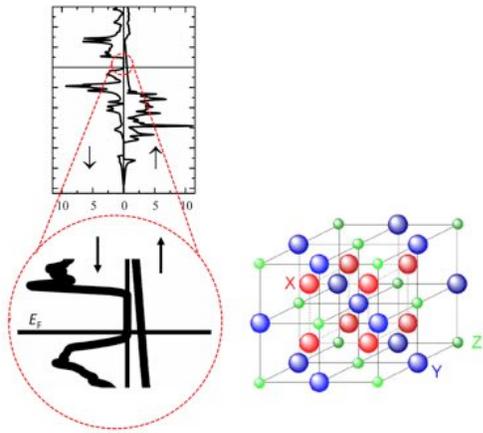

Fig. 1. Minority-spin bandgap [7] and $L2_1$ phase [6] of the full-Heusler alloys.

Regarding (m1.1), in 2011, 74.8% GMR ratio was reported by Sato *et al.* [8] using a junction consisting of $Co_2Fe_{0.4}Mn_{0.6}Si/Ag/Co_2Fe_{0.4}Mn_{0.6}Si$. This is a significant improvement from 41.7% reported by Takahashi *et al.* about 5 months earlier. By using such a GMR junction as a read head, he GMR ratio of approximately 75% with the resistance area product ($RA$) of almost 0.17 $\Omega \cdot \mu m^2$ satisfies the requirement for 2-Tbit/in$^2$ areal density in a hard disk drive (HDD). Figure 2 shows the requirement and recent major efforts towards the Tbit/in$^2$ areal density. It is clear that the Heusler-alloy GMR junctions are the only candidates satisfying the requirement to date. By reflecting on the development over the last 5 years, one can expect the Heusler-alloy GMR junctions can achieve 100% GMR ratios within 3 years. This will satisfy (m1.1) and will lead to device applications as HDD read heads.

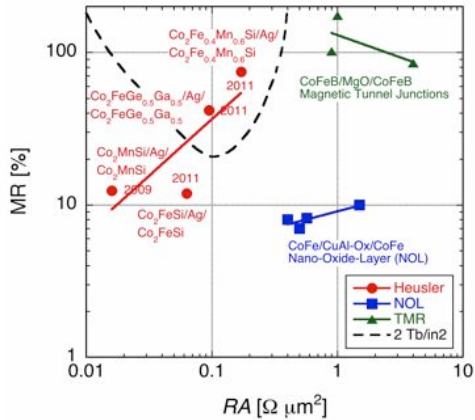

Fig. 2. Requirement for Tbit/in$^2$ HDD read head and recent major results [9].

For (m1.2), Figure 3 shows the development of the TMR ratios using amorphous and MgO barriers with both conventional ferromagnets and Heusler alloys as electrodes. As shown here the largest TMR reported to date is 604% at RT using a magnetic tunnel junction (MTJ) of CoFeB/MgO/CoFeB [10]. In 2005, an MTJ with an epitaxial $L2_1$ $Co_2MnSi$ film has been reported to show very high TMR ratios of 70% at RT [11]. These are the largest TMR ratio obtained in an MTJ with a Heusler alloy film and an Al-O barrier. The TMR is purely induced by the intrinsic spin polarisation of the Heusler electrodes which is different from an MTJ with an oriented MgO barrier, where a TMR ratio of 386% has been achieved at RT (832% at 9 K) for $Co_2FeAl_{0.5}Si_{0.5}$ [12]. The TMR ratio reported here is the highest ever in an MTJ with a Heusler alloy film but with the assistance of coherent tunnelling through an oriented MgO barrier. By taking a moderate extrapolation, one can estimate that 1,000% TMR ratios (m1.2) can be achieved within 10 years time period, *i.e.*, the RT half-metallicity by 2024.

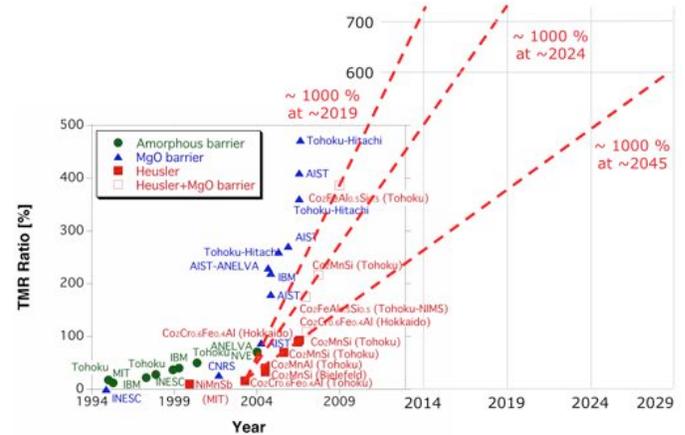

Fig. 3. Recent developments in the TMR ratios.

The other device application expected is to fabricate all Heusler junctions consisting of antiferromagnetic/ferromagnetic/non-magnetic/ferromagnetic Heusler-alloy layers. Such junctions can offer a template to avoid any crystalline disorder at the interfaces as the lattice matching and symmetry can be precisely controlled by atom substitution in these alloy layers. As a first step, Nayak *et al.* reported an antiferromagnetic Heusler alloy of $Mn_2PtGa$ for the first time but at low temperature (below 160K) [13]. One can anticipate RT antiferromagnetism can be demonstrated within 5 years, leading to the all Heusler-alloy junctions in 20 years.

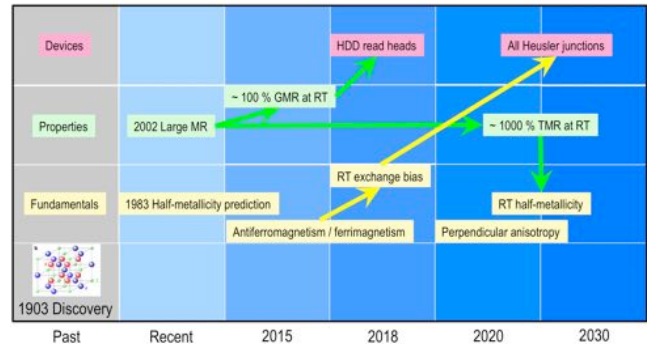

Fig. 4. Roadmap on the Heusler-alloy films.

By summarising the above consideration, one can anticipate a roadmap on the Heusler-alloy films as shown in Fig. 4. The Heusler-alloy films are expected to be used in GMR read heads and sensors within 3 years. These films are also to be combined with antiferromagnetic and/or non-magnetic Heusler-alloy films to form all Heusler junctions. Such junctions may be used in a magnetic random access memory subject to their perpendicular magnetic anisotropy, which is still in the infant stage in research.

## II. OXIDES

Ferromagnetic oxide thin films have been intensively studied for more than last two decades due to their a large variety and tunability of physical properties such as ferro-, ferri-, anti-ferromagnetism, ferroelectricity, superconductivity, optical properties, and colossal magnetoresistance effect [14], [15]. In particular, some of ferromagnetic oxides are predicted as promising candidates of a half-metal and a spin-filter, which directly lead to a large magnetoresistance (MR) as discussed in the previous section. In addition, due to a high compatibility with other oxides and organic materials, establishment of high quality all oxide heterostructure beyond CMOS device are highly expected. In this section, milestones and their associated roadmaps for 3 half-metallic oxide ferromagnets, spinel ferrites (2.1), rutiles (2.2) and perovskites (2.3), are discussed.

### A. Spinel Ferrites

The most commonly studied oxides of Fe is $Fe_3O_4$, which has an inverse spinel structure and a magnetic moment of 4.1 $\mu_B$/f.u. [16]. Among the various spinel-type ferrites, $Fe_3O_4$ is a major conductive oxide at room temperature. The Curie temperature $T_C$ is ~850 K and the characteristic metal-insulator transition point (Verwey temperature) is 123 K. According to a band calculation, half-metallicity has been predicted [18],[19] and spin-resolved photoemission experiments show that $Fe_3O_4$ exhibits spin polarisation of up to −80% [20]. Very high spin polarisation has also been suggested by the measurement of an MR ratio of over 500% through a nano-contact [21]. Epitaxial $Fe_3O_4$ films have been grown by various techniques, including molecular beam epitaxy (MBE) under an oxygen atmosphere, magnetron sputtering and laser ablation [20]. By replacing one of the Fe ions with a divalent metal ion, $e.g.$, Mn, Co, Ni etc., a ferrite can be formed [20]. Pénicaud has predicted half-metallicity in Mn, Co and Ni ferrites [22] although the bulk materials are insulators except $Fe_3O_4$. In particular, $NiFe_2O_4$ shows a bandgap in the majority band, indicating that this compound can become an insulator or semi-metallic half-metal. The discrepancy of the bandgap structure between *ab initio* calculation results and experimental results suggests that the treatment of electron correlation is significant.

Some ferrites are expected as a good candidate of a spin-filter because of their ferromagnetic insulator properties and high $T_C$. The spin-filtering device consists of a ferromagnetic insulator layer sandwiched between a non-magnetic metallic (NMM) layer and a ferromagnetic metallic (FMM) layer (or a superconductive layer). Due to the exchange splitting of the energy levels in the conduction band of the ferromagnetic insulator, the effective barrier height for the up-spin electron differs from that for the down-spin one, leading to a large difference in the tunnelling probabilities between the two spin orientations. Therefore, ideally, an almost perfectly spin-polarized current is generated and this results in an infinite MR if a ferromagnetic insulator with a large exchange splitting is used. Here, MR ratio is defined as $2P_{SF}P/(1 - P_{SF}P)$, where $P_{SF}$ is the spin filtering efficiency [= $(I_{up} - I_{down})/(I_{up} + I_{down})$, $I_{up(down)} \propto \exp(-d \cdot \phi_{up(down)}^{1/2})$, where $I$ is the tunnelling current, $d$ is the thickness of the spin-filter, and $\phi$ is the effective barrier height] and $P$ is the spin polarisation of the FMM layer. RT spin-filtering effect has been demonstrated using $CoFe_2O_4$-based spin-filter devices [23],[24]. However $|P_{SF}|$ at RT is below 5%.

Related to Section 4, perpendicular magnetization behaviour with a high uniaxial magnetic anisotropy of $K_u = 1.47 \times 10^6$ J/m$^3$ in $CoFe_2O_4$ ferrite [25] has been reported. In addition to the ferromagnetic spinel ferrites, nonmagnetic spinel, $MgAl_2O_4$ has also attracted much attention as a new spintronics material because an ultrathin $MgAl_2O_4$ layer shows coherent tunnelling properties (symmetry selective tunnelling) and high MR ratios like an MgO tunnel barrier. Using an epitaxial CoFe/$MgAl_2O_4$ (with cation-site disordered)/CoFe structure, an MR ratio of more than 300% at room temperature was reported [26].

Towards the magnetic ferrites as a spintronic material, the following milestones have been recognised:

(m2.1.1) Half-metallic behaviour and high MR by improving microstructure and control of interface states;

(m2.1.2) High spin-filtering effects at RT by reducing structural and chemical defects;

(m2.1.3) Tuning of perpendicular magnetic anisotropy;

(m2.1.4) Development of new non-magnetic spinel-based materials to tune the transport properties and the coherent tunnelling effect.

Regarding (m2.1.1) and (m2.1.2), ferrite films with a very high quality crystalline structure, $i.e.$, without any crystal imperfections such as anti-phase boundaries (APBs), atomic site disorder and dislocations, are necessary to obtain high saturation magnetisation, high squareness of the hysteresis loops and high $T_C$. The presence of APBs within a ferrite film, for instance, significantly degrades the saturation magnetisation under a high magnetic field and the remanence. It also increases the resistivity of the film since the APBs induce the electron scattering centre. Consequently, high quality films are indispensable to achievement of stable half-metallic characteristics and a spin-filtering effect at RT. In addition, realisation of a perfect and an abrupt ferrite/non-magnetic interface is required to preserve high effective spin polarisation at the interface states. Therefore, establishment of the growth method and procedures for high quality ferrite films, as well as a high quality interface with ferro- and non-magnetic metallic layers are strongly desired. The development of an advanced growth process will lead to RT half-metallicity using ferrite family materials such as $Fe_3O_4$, γ-$Fe_2O_3$, $CoFe_2O_4$, $NiFe_2O_4$, $MnFe_2O_4$, and $ZnFe_2O_4$.

The milestone of (m2.1.3) is important to ensure the high thermal stability for nano-scale structures using $CoFe_2O_4$-based ferrites for future spin-filtering devices and other spintronics use at RT. Especially, strong perpendicular magnetic anisotropy in a very thin region (below several nm) is desirable to control the tunnelling resistance for device applications.

For (m2.1.4), providing the new non-magnetic tunnel barrier is now considered as an important issue to establish novel spintronic hetero-structures since only a limited tunnel barrier material ($Al_2O_3$ and MgO) are currently available to obtain high RT MR ratios. Especially, the ability to tune physical properties is required to achieve higher performance, multi-functionality and better compatibility to ferromagnetic electrodes. For instance, MR enhancement by crystalline barrier (coherent tunnelling), a perfect lattice matching (lattice constant tuning), a low tunnelling resistance (barrier height tuning), and applicability of high electric fields to a ferromagnetic layer facing the barrier (dielectric constant tuning) are presumably possible in spinel-based nonmagnetic barrier with tailored compositions.

In summary, one can propose a roadmap on spinel ferrite films as shown in Fig. 5. Using spinel ferrite based MTJs consisting of ferrite/non-magnetic (NM) barrier/ferrite (or FMM) structure, >100% RT TMR (corresponding $|P|$ is ~0.7 according to the Julliere model) is expected within 10 years through development of high quality spinel ferrite thin films and selection of a proper NM barrier. Further improvement of an MTJ structure and suppression of a rapid TMR reduction with increasing temperature will lead to giant TMR over 1000% (corresponding $|P|$ is ~0.9) within 25 years.

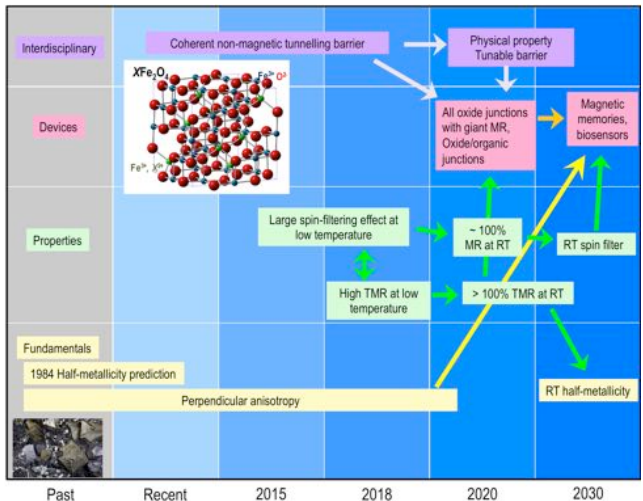
Fig. 5. Roadmap on the ferrite films.

To construct spin-filtering devices, one can use the techniques for the MTJ fabrication; a typical stacking structure is NMM/ferrite spin-filter/NM barrier/FMM, where the NM barrier is used to weaken the exchange coupling between the ferrite and FMM layers. Recently a higher $P$ of 8% at RT (MR ~ 6%) has been demonstrated using an epitaxial $Pt/CoFe_2O_4/Al_2O_3/Co$ nano-contact junction [27]. Thus, improvement of the junction structure as well as the ferrite film quality can enhance MR ratio. >100% RT MR ratio due to the spin-filtering effect is expected within 10 years by reducing structural and chemical defects in spin-filter junctions.

Using new NM barriers, one can highly expect a giant TMR ratio exceeding 500% at RT within 5 years. Furthermore, the tuning of physical properties will be achieved by searching for new candidate barrier materials within 10 years.

### B. Rutiles

Using Andreev reflection, $CrO_2$ has been proven to show a half-metallic nature at low temperature as suggested by *ab initio* calculations [16],[17]. High spin polarisation of 90% has been confirmed at low temperature using point-contact Andreev reflection method [18],[19] and high powder magnetoresistance has been reported [20]. However, RT half-metallicity has not been demonstrated yet. $CrO_2$ has a tetragonal unit cell with a magnetic moment of 2.03 $\mu_B$/f.u. at 0 K [21]. The ferromagnetism of $CrO_2$ appears below 391 K [22]. Above this temperature another phase of $Cr_2O_3$ is known to show antiferromagnetism, which is the major cause of the reduction of the half-metallicity. Highly-ordered $CrO_2$ films are predominantly grown by chemical vapour deposition [23]. However, obtaining the $CrO_2$ single phase as a thin film is not easy, and thus MR properties steeply decrease below RT.

In order to utilise the rutiles in a spintronic device, the following milestones have been identified:

(m2.2.1) Development of a high quality $CrO_2$ thin film with a single rutile phase and achievement of a clean interface structure with tunnel junctions;

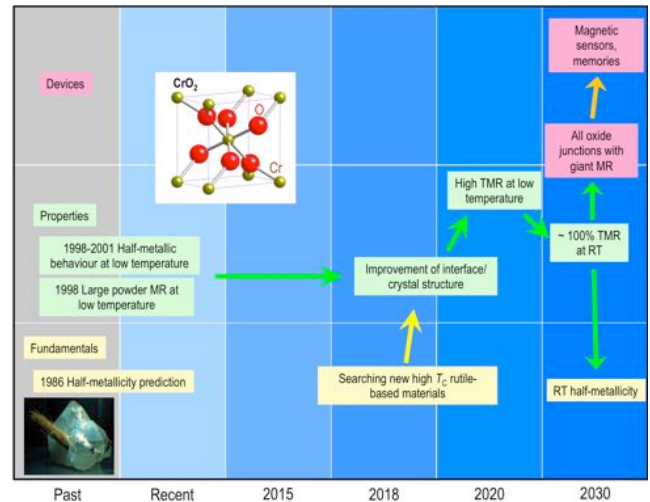
Fig. 6. Roadmap on the rutile films.

(m2.2.2) Search for new rutile-based materials with higher $T_C$ and robust half-metallicity by tailoring their composition.

Regarding (m2.2.1), the undesirable reduction in MR ratio below $T_C$ could be suppressed by the improvement of the crystal structure and the interface state. Optimisation of an epitaxial growth process for a single rutile phase and use of a suitable non-magnetic barrier which does not invade the interface of $CrO_2$ will be effective. In addition, the elimination of the nonmagnetic $Cr_2O_3$ phase, which generally forms on the surface of the $CrO_2$ film, using sophisticated deposition and treatment processes will enhance the magnetic and half-metallic properties.

For (m2.2.2), to obtain a more stable half-metallic phase with high $T_C$, doping of other elements to $CrO_2$ or searching ternary or quaternary rutile-based ferromagnetic materials would be necessary. Such a new composition and a new material will lead to stable half-metallic properties and higher MR at RT.

In summary, one can anticipate a roadmap on the half-metallic rutile films as shown in Fig. 6. Obtaining epitaxial thin films with a single $CrO_2$ phase will lead to observation of RT TMR ratios within 10 years. To demonstrate high TMR ratios (>100%) at RT is still challenging. Searching new rutile type ferromagnetic oxides and a sophisticated MTJ structure might yield a technological breakthrough toward a higher TMR ratio in the future.

*C. Perovskites*

Perovskites, such as $(La,Sr)MnO_3$, exhibit both strong ferromagnetism and metallic conductivity with partial substitution of $La^{+3}$ ions with $2^+$ ions such as Ca, Ba, Sr, Pb and Cd [28],[29]. Since only one spin band exists at $E_F$ in these films, 100% spin polarisation can be achieved. Using these materials instead of a conventional ferromagnet, a very high MR of ~150% at RT has been observed [30]. This is known as colossal magnetoresistance (CMR). Using Mn-perovskite thin films and $SrTiO_3$ oxide tunnel barrier, a TMR ratio of up to 1850% has been reported but only below $T_C$ [31]. CMR can be induced either by breaking the insulating symmetry of $Mn^{3+}$ and $Mn^{4+}$ alternating chains or by suppressing spin fluctuation near $T_C$. Even so, it is unlikely to achieve the RT half-metallicity in the conceivable future.

Much effort has been spent to search for new high $T_C$ perovskites for a RT half-metallicity. The family of double perovskites with a chemical composition of $A_2BB'O_6$ (A is an alkaline earth or rare earth ion, B and B' are transition metal ions), has been focused for more than 15 years since some of the double perovskites exhibit high $T_C$ above RT and half-metallic band structures [32]. $Sr_2FeMoO_6$ (SFMO) has high $T_C$ of 420 K and has been predicted to be a half-metal [33], indicating the double perovskites are a promising oxide family for high MR at RT. At low temperature, high $P \sim$ 80% in a SFMO film has been demonstrated using a $Co/SrTiO_3/SFMO$ MTJ. Much higher $T_C$ of 635 K is reported in $Sr_2CrReO_6$ [34].

Recently 2-dimensional electron gas (2DEG) at the interface of a nonmagnetic perovskite hetero-structure consisting of $LaAlO_3/SrTiO_3$ has been investigated intensively due to a high mobility in the 2DEG. Highly efficient spin transport in the 2DEG could be usable to establish the new type spin transistors in the future.

The following milestones have been established towards the perovskites as a spintronic material:

(m2.3.1) Search for new perovskite-based materials with $T_C$ > RT;

(m2.3.2) Development of a high MR at RT.

Regarding (m2.3.1), the double perovskites with $A_2FeMoO_6$ or $A_2FeReO_6$ series are promising due to their high $T_C$. However, high MR using an MTJ structure has not been achieved since there are some considerable obstacles against (m2.3.2); (1) site disorder of magnetic ions deteriorates the magnetic properties and the spin polarisation, and (2) their high reactivity to water, which restricts use of common microfabrication techniques.

In order to overcome these obstacles, improvement of film quality and preparation of a clean interface are necessary to achieve high MR ratios at RT. Especially, specific microfabrication method should be newly developed to reduce the damage during the processes. In addition, a new barrier material that matches with the perovskites will be needed to compose a high quality perovskite-based MTJ.

In summary, one can expect a roadmap on the perovskite films as shown in Fig. 7. RT TMR ratios will be obtained using MTJs with a high $T_C$ perovskite layer within 5 years. >100% TMR at RT will be expected in the future after demonstration of high TMR ratios at low temperatures.

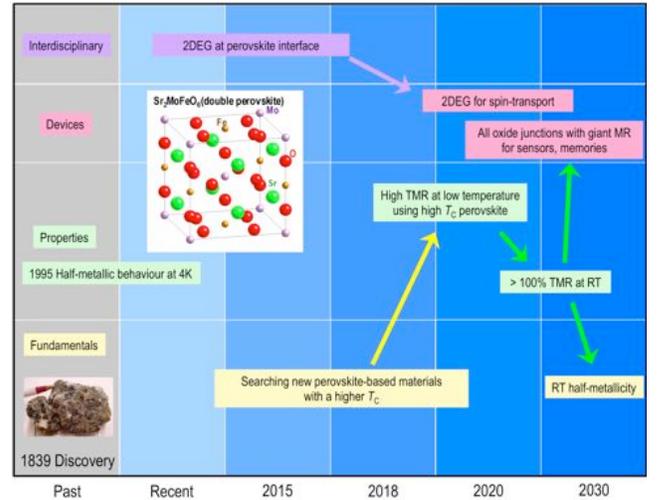

Fig. 7. Roadmap on the perovskite films.

III. DILUTE MAGNETIC SEMICONDUCTORS

Unlike metals, semiconductors have relatively low carrier density that can be drastically changed by doping, electrical gates, or photo-excitations, to control their transport and optical properties. This versatility makes them the materials of choice for information processing and charge-based electronics. In magnetically-doped semiconductors, such as (Cd,Mn)Te, (In,Mn)As, or (Ga,Mn)As, these changes of carrier density also enable novel opportunities to control magnetic properties and lead to applications that are not available or ineffective with ferromagnetic metals [35]. For example, a carrier-mediated magnetism in semiconductors offers tunable control of the exchange interaction between carriers and magnetic impurities. The onset of ferromagnetism and the corresponding change in the $T_C$ can be achieved by increasing the carrier density using an applied electric field, photo-excitations, or even heating. Two milestones for the research on novel magnetic semiconductors are identified:

(m3.1) search for tunable ferromagnetism in semiconductors with $T_C$ > RT.

(m3.2) demonstrating RT devices that are not limited to magnetoresistive effects.

Considering (m3.1), despite numerous reports for $T_C$ > 300 K in many semiconductors, a reliable RT ferromagnetic semiconductor remains elusive [36],[37]. However, even the existing low-$T_C$ magnetic semiconductors have provided demonstrations of novel magnetic effects and ideas that have subsequently been also transferred to ferromagnetic metals, for example, electric-field modulation of coercivity and magnetocrystalline anisotropy at RT [37].

An early work on ferromagnetic semiconductors dates back to $CrB_3$ in 1960 [38]. Typically studied were concentrated magnetic semiconductors, having a large fraction of magnetic elements that form a periodic array in the crystal structure. Important examples are Eu-based materials in which the solid-state spin-filtering effect was demonstrated for the first time [39]. However, complicated growth and modest $T_C$ (up to ~150 K) limited these materials to fundamental research. Starting with mid-1970s, the dilute magnetic semiconductors (DMS), alloys of nonmagnetic semiconductor and magnetic elements (typically, Mn) [40], became intensely explored first in II-IV, and later in III-V nonmagnetic hosts. In II-VIs $Mn^{2+}$ is isovalent with group II providing only spin doping, but not carriers and thus making robust ferromagnetism elusive. In III-Vs Mn yields both spin and carrier doping, but low-Mn solubility limit complicates their growth and can lead to an extrinsic magnetic response due to nanoscale clustering of metallic inclusions. This complex dual role of Mn doping in III-Vs possess both: (i) challenges to establish universal behavior among different nonmagnetic III-V hosts. (Ga,Mn)N predicted to have $T_C > 300$ K [41], but shown to only have $T_C \sim 10$ K [42], (ii) makes the *ab initio* studies less reliable, requiring careful considerations of secondary phases and magnetic nano-clustering – a source of many reports for an apparent high-$T_C$ in DMS.

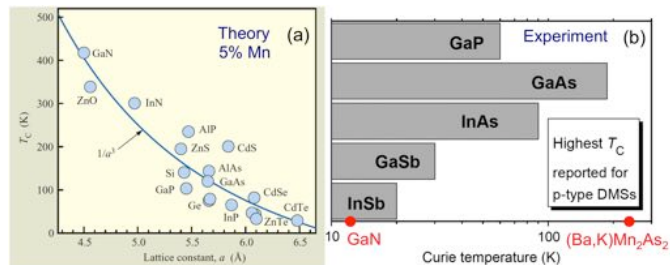

Fig. 8. (a) Theoretical predictions for $T_C$ in DMS [41], adapted from Ref. [45]. (b) Reliable highest experimental $T_C$ reported for Mn-doped DMS, adapted from Ref. [36].

An important breakthrough came with the growth of III-V DMS: (In,Mn)As in 1989 and (Ga,Mn)As in 1996 [43],[44], responsible for demonstrating tunable $T_C$, coercivity, magnetocrystalline anisotropy, as well as the discovery of tunnelling anisotropic magnetoresistance [37]. However, even if the low-Mn solubility is overcome (maximum ~ 10%), the upper $T_C$ limit is given MnAs with $T_C \sim 330$ K. This suggests that (Ga,Mn)As, with the current record $T_C \sim 190$ K [41], is not a viable candidate for RT ferromagnetism in DMS. Influential mean-field calculations [39] for DMS with 5% Mn in Fig. 8(a) show a strong correlation with an inverse unit cell volume [45]. However, *ab initio* studies reveal a more complex, material-dependent situation [46].

Instead of III-V compounds, more promising are recently discovered II-II-V DMS [47]. They are isostructural to both 122 class of high-temperature Fe-based superconductors and antiferromagnetic $BaMn_2As_2$, offering intriguing possibilities to study their multilayers with different types of ordering. In $(Ba,K)(Zn,Mn)_2As_2$ with an independent carrier (K replacing Ba) and spin doping (Mn replacing Zn), some of the previous limitations are overcome: the absence of carriers in II-VIs and the low-Mn solubility in III-Vs. With 30 % K and 15 % Mn doping, their $T_C \sim 230$ K [48] exceeds the value in (Ga,Mn)As. Selected highest reliable experimental $T_C$ reported for Mn-doped DMS are shown in Fig. 8(b). Circles are given for GaN which has about 30 times smaller $T_C$, than predicted in panel (a), and $(Ba,K)Zn_2As_2$, a current record for DMS. *Ab initio* studies predict a further increase in $T_C$ [49]. We expect that tunable RT carrier-mediated ferromagnetism will be realized in II-II-V DMS within 5 years.

(m3.2) While DMS are often viewed as the materials for multifunctional devices to seamlessly integrate nonvolatile memory and logic [35], other device opportunities could be more viable. In fact, DMS-based optical isolators [50],[51] were already commercialized by Tokin Corporation [52]. Such devices, relying on large magneto-optical effects (Faraday and Kerr) that are proportional to the giant Zeeman splitting in DMS, are used to prevent feedback into laser cavities and provide one-way transmission of light. Even without demonstrating $T_C >$ RT, enhancing RT Zeeman splitting is important for DMS (exceeding a large $g$-factor ~ 50 for InSb).

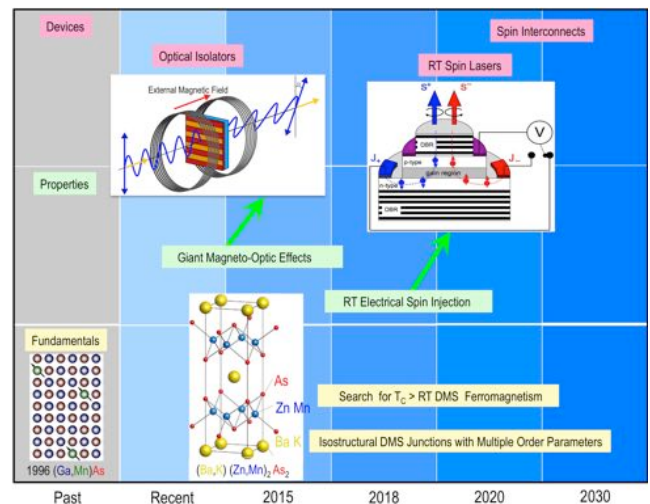

Fig. 9. Roadmap on dilute magnetic semiconductors.

Spin-lasers [53],[54] are another example of devices not limited to MR effects. They can outperform [55],[56] conventional lasers with injected spin-unpolarized carriers. For spin-lasers electrical spin injection is desirable, currently limited up to ~ 230 K [57]. $T_C >$ RT in DMS would be beneficial for such spin-lasers, both as an efficient spin injector and possibly a tunable active region that could alter the laser operation through tunable exchange interaction. To remove the need for an applied $B$-field, perpendicular anisotropy of the spin injector is suitable. We expect RT electrical spin injection in spin-lasers by 2020. It is important to critically assess if extrinsic $T_C >$ RT in DMS, from magnetic metallic nano-inclusions and secondary phases [having GaAs+MnAs, rather than (Ga,Mn)As, a true DMS] is a viable path for RT spintronic devices. RT magnetoamplification was demonstrated in (In,Mn)As-based magnetic bipolar transistor, operating above the $T_C$ < 100 K of a single-phase (In,Mn)As [58]. Another test for useful extrinsic (multiphase) DMS is a robust RT electrical spin injection. A road map for DMS is given in Fig. 9.

## IV. Perpendicularly Anisotropic Ferromagnets

A perpendicularly magnetised system is currently an important building block in spintronic devices since it enables us to shrink the size of memory bits and to reduce the electric current density required for spin-transfer switching. There are several ways to obtain perpendicular magnetic anisotropy in a thin film. To use an ordered alloy showing high magnetocrystalline anisotropy is one possible way. If its easy magnetisation axis is aligned along the normal direction to the film plane and the magnetocrystalline anisotropy field overcomes the demagnetisation field, the film shows the perpendicular magnetisation. Another way is to use the interface magnetic anisotropy between a ferromagnetic layer and a non-magnetic layer. In addition, multilayered structures are useful to obtain perpendicular magnetisation.

Towards the perpendicularly anisotropic ferromagnet as a spintronic material, the following milestones have been established:

(m4.1) High thermal stability of perpendicular magnetisation;
(m4.2) Structural stability against the thermal process;
(m4.3) Demonstration of high spin-polarisation;
(m4.4) Reduction of the magnetic damping constant.

(m4.1) means the stability of magnetisation at a nanometer scale overcoming the magnetisation fluctuation due to the thermal energy. Considering several thermal treatments in device fabrication processes, (m4.2) should be satisfied. (m4.3) is a key determining the performance of MTJ and GMR devices. In terms of spin-transfer torque (STT) magnetisation switching, as indicated in (m4.4), the magnetic damping should be small to reduce the electric current density for switching.

An $L1_0$-ordered structure exists in the thermodynamically stable phase and is composed of the alternative stacking of two kinds of atomic planes along the $c$-axis. Thus, $L1_0$-ordered alloys, such as FePt, FePd, CoPt, MnAl and MnGa, exhibit uniaxial magnetic anisotropy along the $c$-axis direction. When one aligns the $c$-axis of $L1_0$-ordered structure in the normal direction to the film plane, a perpendicular magnetic anisotropy is obtained. Since the $L1_0$-ordered structure is thermally stable, $L1_0$-ordered alloys have an advantage from the viewpoint of (m4.2). Among the $L1_0$-ordered alloys, $L1_0$-FePt shows the largest uniaxial anisotropy ($K_u$) of $7 \times 10^6$ J/m$^3$ [59], which leads to the excellent thermal stability of magnetisation at a reduced dimension, e.g., 4 nm diameter in $L1_0$-FePt nano-particles. This property satisfies (m4.1). Thanks to its perpendicular magnetisation for FePt (001) films, $L1_0$-FePt has been regarded as an ideal material for perpendicular recording media in a HDD. In addition, the spin polarisation of FePt is theoretically predicted to be approximately 70% [60], which is a good characteristic for a spintronic material. $L1_0$-ordered FePt films have already been implemented in both MTJ [61] and GMR [60] junctions. In the case of GMR nano-pillars consisting of two FePt layers separated by nonmagnetic Au, the STT phenomena have been examined systematically by tuning the crystalline order of the FePt layer [60]. However, the observed TMR and GMR ratios are still low for $L1_0$-FePt.

Another important issue is that the major $L1_0$-ordered alloys contain the heavy transition metals such as Pt. The Pt atom shows strong spin-orbit coupling, which leads to the significant enhancement of magnetisation damping. This feature is an opposite trend to (m4.4). $L1_0$-FePd exhibits a large $K_u$ and rather smaller damping constant compared with that of $L1_0$-FePt, probably because Pd is lighter element than Pt [62]. However, the usage of such noble metals as Pt and Pd is not suitable from the viewpoint of element strategic trend. Considering these recent demands, a new kind of $L1_0$-alloy is eagerly desired, which possesses a large $K_u$ and a small damping constant. One of the candidates is $L1_0$-FeNi. Since a paper reported that an $L1_0$-FeNi bulk alloy exhibited high uniaxial magnetic anisotropy of $K_u = 1.3 \times 10^6$ J/m$^3$ [63], $L1_0$-FeNi is a future material having a possibility to substitute high $K_u$ materials containing the noble metals and rare earths. Kojima et al. reported the preparation of $L1_0$-FeNi thin films with relatively high $K_u$ of $0.7 \times 10^6$ J/m$^3$ [64], and also the small damping constant has been reported in $L1_0$-FeNi [65].

Another candidate material showing perpendicular magnetisation is a Mn-based alloy system, such as $L1_0$-MnAl. Recently, epitaxial Mn-Ga films including $L1_0$ and $D0_{22}$ ordered phases have also been found to exhibit strong perpendicular magnetic anisotropy ($K_u = 1.2$-$1.5 \times 10^6$ J/m$^3$) with small saturation magnetisation ($M_S = 250$-$500$ emu/cm$^3$) and small magnetic damping ($\alpha = 0.0075$-$0.015$) at RT [66],[67]. Moreover, it has been found that $D0_{22}$-Mn$_3$Ge epitaxial films exhibited $K_u$ of $0.91 \times 10^6$ J/m$^3$ [68] and $1.18 \times 10^6$ J/m$^3$ [69]. These Mn-based alloy systems can also be used as a perpendicular magnetised layer for STT-application because the ab-initio calculations predicted the high spin polarisation of 88% for Mn$_3$Ga [70] and a half-metallic band dispersion for Mn$_3$Ge that leads a high TMR like Fe/MgO-MTJs [71],[72]. However, the observed TMR ratios are also still low for $L1_0$- and $D0_{22}$-Mn-Ga [73]. Experimental realisation of high spin polarisation is essential for all the ordered alloys to achieve (m4.3).

Multilayered structures, such as Co/Pt, Co/Pd, Co/Ni and so on, also show perpendicular magnetisation. The main origins for perpendicular magnetic anisotropy in the multilayered structures are as follows: (i) breaking the crystal symmetry at the interface, which leads to the interface magnetocrystalline anisotropy, (ii) the effect of magnetostriction due to the interface between different atomic planes, or (iii) interface alloying. Although the multilayered films show high magnetic anisotropy, we need to consider the stability of the layered structure against a thermal process. In some cases, the high temperature annealing degrades the layered structure and its magnetic properties, which should be improved for (m4.2). Mangin *et al.* [74], and Meng and Wang [75] also demonstrated the STT switching in CPP-GMR nano-pillars with perpendicularly magnetised Co/Ni and Co/Pt multilayers, respectively. As in the case of the ordered alloys, however, increasing MR effect and lowering magnetization damping are inevitable issues for the multilayered structures to achieve (m4.3) and (m4.4). To explore the adequate materials combination is one of the ways for the multilayered structure to solve the current problems.

One of the new types of multilayering films is an artificial superlattice grown by using nearly mono-atomic-layer alternation of Co and Pt or Pd. Such ultrathin superlattice films had an annealing stability higher than that of conventional multilayering films [76].

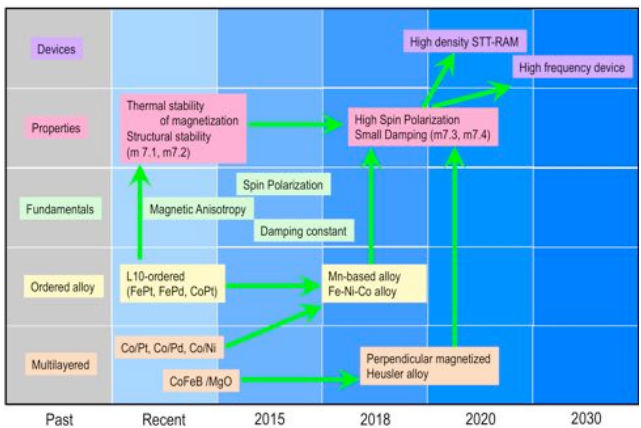
Fig. 10. Roadmap on the perpendicularly anisotropic films.

It has also been reported that the CoFeB/MgO junction shows perpendicular magnetic anisotropy [77]. The perpendicular magnetisation components of the CoFeB are induced at the MgO interface, which originates from the interface magnetic anisotropy. The perpendicularly magnetised CoFeB/MgO layers have a significant advantage because MgO-based tunnel junctions show high TMR ratio. Actually, it has also been demonstrated that a CoFeB/MgO/CoFeB stack with perpendicular magnetisation shows the TMR ratio over 120% and the low STT switching current of 49 µA at a 40-nm-diameter junction. This is a promising candidate as a building block for the MRAM cell. However, because the interfacial magnetic anisotropy constant is not large enough and a thin ferromagnetic layer is required to exploit the interface effect, the small volume of the magnetic layer may give rise to the thermal instability of magnetisation in a deeper sub-nanometer region. (m4.1) is an important step for the perpendicular anisotropic ferromagnets using the interface magnetic anisotropy. Also, perpendicularly magnetised Heusler alloy layers, where interface magnetic anisotropy is used, are attracting attention as an alternative perpendicularly magnetised system, which may lead to high spin polarisation (m4.3) and a low damping constant (m4.4). Recently, perpendicular magnetization and the TMR ratio of 132% at room temperature have been demonstrated using an ultra-thin $Co_2FeAl$ Heusler alloy/MgO/CoFeB MTJ [78].

## V. OVERVIEW

In this roadmap, we have identified two key properties to develop new (and/or improved) spintronic devices. The first one is the half-metallicity at room temperature (RT), which can be achieved by clearing milestones to realise large MR and resulting large spin polarisation. The second one is the perpendicular anisotropy in nano-scale devices at RT. This is based on milestones, including large perpendicular magnetic anisotropy and small damping constant. Such development is expected to be achieved not only by the development of these alloys but also by the fundamental understanding on these properties using a well-studied test system, *i.e.*, zincblendes. As summarised in Fig. 11, we anticipate these materials investigated here to realise all Heusler and all oxides junctions. These can be implemented in next-generation MRAM and high-frequency devices within 35 years.

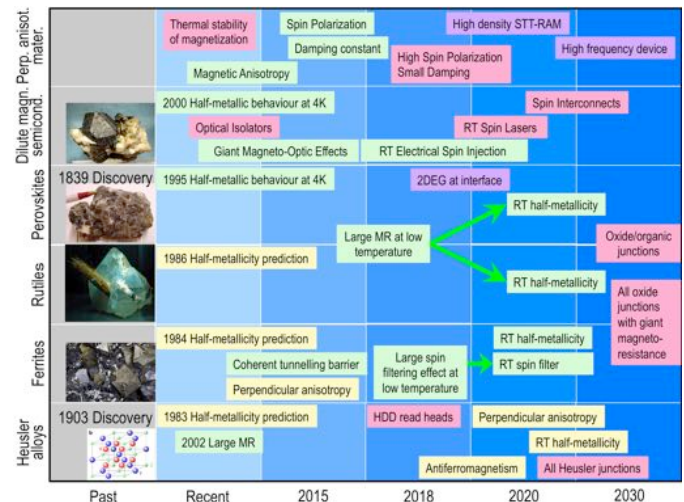
Fig. 11. Roadmap for magnetic materials.

ACKNOWLEDGMENT

The authors wish to thank the Technical Committee of the IEEE Magnetics Society, who initiated this roadmap as the first exercise of this kind. AH and TS appreciate financial support by PRESTO-JST. AH also appreciates the financial support by EPSRC (EP/K03278X/1 and EP/M02458X/1) and EU-FP7 (NMP3-SL-2013-604398). SM thanks to a NEDO Development of an Infrastructure for Normally-off Computing Technology project and ASPIMATT (JST). IZ was supported by US ONR N000141310754, NSF ECCS-1508873, NSF ECCS-1102092 and NSF DMR-1124601.

**Atsufumi Hirohata** (M'01–SM'10) was born in Tokyo in 1971. He received his BSc and MSc in Physics from Keio University in Japan in 1995 and 1997, respectively. He then received his PhD in Physics from the University of Cambridge in the U.K. His major fields of study are spintronic devices and magnetic materials.

Hirohata was a Post-Doctoral Research Associate at the University of Cambridge and the Massachusetts Institute of Technology. He then served as a Researcher at Tohoku University and RIKEN in Japan. He became a Lecturer at the University of York in the U.K. in 2007 and was promoted to be a Reader in 2011, followed by a Personal Chair appointment since 2014. He has edited *Epitaxial Ferromagnetic Films and Spintronic Applications* (Karela, Research Signpost, 2009) and *Heusler Alloys* (Berlin, Springer, 2015). Current research interests are spin injection in ferromagnet/semiconductor hybrid structures, lateral spin-valve devices, magnetic tunnel junctions and Heusler alloys.

Prof. Hirohata is a member of the American Physical Society, Materials Research Society, Institute of Physics, Magnetics Society of Japan, Physical Society of Japan and Japan Society of Applied Physics. In the IEEE Magnetics Society, he served as a member of the Administrative Committee between 2012 and 2014, and is a member of the Technical Committee since 2010.

**Hiroaki Sukegawa** received his MEng degree in materials science from Tohoku University, Sendai, Japan in 2004, and his PhD degree in Materials Science from Tohoku University, in 2007. He became a researcher at National Institute for Materials Science in 2007. He is currently a senior researcher of Magnetic Materials Unit, National Institute for Materials Science. His research interests include magnetic thin films and spintronics devices.

**Hideto Yanagihara** received his BSc and MSc in Materials Science from Keio University in Japan in 1993 and 1995, respectively. He then received his PhD in Applied Physics from the University of Tsukuba in Japan.

Yanagihara was a Post-Doctoral Research Associate at the University of Tsukuba and the University of Illinois at Urbana-Champaign. Current research interests are magnetic thin films and oxides.

Prof. Yanagihara is a member of the American Physical Society, Japan Society of Applied Physics, Magnetics Society of Japan and Physical Society of Japan.

**Igor Žutić** was born in Zagreb, Croatia in 1967. He received his BSc and PhD in Physics from the University of Zagreb, Croatia in 1992 and the University of Minnesota in 1998, respectively.

He was a postdoc at the University of Maryland and at the Naval Research Lab. In 2005, he joined the University at Buffalo, The State University of New York, as an assistant professor and was promoted to an associate professor in 2009 and a full professor in 2013. Following the success of Spintronics 2001: International Conference on Novel Aspects of Spin-Polarized Transport and Spin Dynamics, at Washington, DC, which he proposed and chaired, he was invited to write a comprehensive review, Spintronics: Fundamentals and Applications, for the Reviews of Modern Physics. The review written with Jaroslav Fabian and Sankar Das Sarma is currently among the most cited articles on spintronics and magnetism. With Evgeny Tsymbal he co-edited Handbook of Spin Transport and Magnetism (Chapman and Hall/CRC Press, New York, 2011). His research interests include superconductivity, magnetism, and spintronic devices.

Dr Žutić is a member of the American Physical Society and since 2013 of the Technical Committee in the IEEE Magnetics Society. He is a recipient of the 2006 National Science Foundation CAREER Award, 2005 National Research Council/American Society for Engineering Education Postdoctoral Research Award, and the National Research Council Fellowship (2003–2005).

**Takeshi Seki** was born in Shizuoka, Japan in 1980. He received his BEng, MEng and PhD in Materials Science from Tohoku University in Japan in 2002, 2003 and 2006, respectively. His major field of study is the materials development for spintronic devices.

Seki was a Post-Doctoral Researcher at Tohoku University and Osaka University in Japan. He then became an assistant professor at Tohoku University in 2010. Current research interests are spin transfer phenomena, magnetization dynamics in a nanosized region, and magnetization reversal mechanism.

**Shigemi Mizukami** was born in Sendai, Japan in 1973. He received his BSc and MSc in Applied Physics from Tohoku University in Japan in 1996 and 1998, respectively. He then received his Ph. D in Applied Physics from the Tohoku University. His major fields of study are spintronic devices, high frequency magnetism, and magnetic materials.

Mizukami was a Research Associate at the Nihon University and promoted to be a Lecturer in 2005. He then became an Assistant Professor at the Tohoku University in Japan in 2008, and then promoted to be an Associate Professor in 2011, and also to be a Professor in 2014. He was one of guest editors of special issues: *Advancement in Heusler compounds and other spintronics material designs and applications* (*J. Phys. D: Appl. Phys.*, 2015). Current research interests are ultra-high frequency magnetization dynamics, low damping Heusler materials, perpendicular magnetic tunnel junctions based on Mn-based tetragonal Heusler-like alloys.

Prof. Mizukami is a member of Magnetics Society of Japan, Physical Society of Japan, Japan Society of Applied Physics, and the Japan Institute of Metals and Materials.

**Dr. Raja Swaminathan** is an IEEE senior member and is a package architect at Intel for next generation server, client and SOC products. His primary expertise is on delivering integrated HVM friendly package architectures with optimized electrical, mechanical, thermal solutions. He is also an expert in magnetic materials synthesis, structure, property characterizations and has seminal papers in this field. He is an ITRS author and iNEMI technical WG chair on packaging and design. He has also served on IEEE micro-electronics and magnetics technical committees. He has 13 patents and 18 peer reviewed publications and holds a Ph.D in Materials Science and Engineering from Carnegie Mellon University.